\documentclass{article}

\usepackage{PRIMEarxiv}

\usepackage[utf8]{inputenc} % allow utf-8 input
\usepackage[T1]{fontenc}    % use 8-bit T1 fonts
\usepackage{hyperref}       % hyperlinks
\usepackage{url}            % simple URL typesetting
\usepackage{booktabs}       % professional-quality tables
\usepackage{amsfonts}       % blackboard math symbols
\usepackage{nicefrac}       % compact symbols for 1/2, etc.
\usepackage{microtype}      % microtypography
\usepackage{lipsum}
\usepackage{fancyhdr}       % header
\usepackage{graphicx}       % graphics
\graphicspath{{media/}}     % organize your images and other figures under media/ folder
\usepackage{siunitx}

\title{HUVECs-encapsulation via Millimeter-sized Alginate Droplets
\thanks{\textit{\underline{HUVECs-encapsulation via Millimeter-sized Alginate Droplets}}: 
\textbf{Tran K, et al.,  DOI:000000/11111.}}
}

\author
{Khanh Tran$^{1}$;
Brenda A.A.B. Ametepe$^{1}$; 
Erika Gomez$^{1}$; \\ Daniel Ramos$^{1}$; Clare Kim$^{1}$;
Ga-Young Kelly Suh$^{1}$;\\ Siavash Ahrar$^{1, \dag}$; Perla Ayala $^{1, \dag}$
\\
\\\normalsize{$^{1}$ Department of Biomedical Engineering, CSU Long Beach, CA, USA}\\
\normalsize{ \dag  Co-corresponding authors.}\\
}

\begin{document}
\maketitle

%The next command sets up an environment for the abstract to your paper.
\newenvironment{sciabstract}{%
\begin{quote}}
{\end{quote}}

\begin{sciabstract}
\begin{center}
\textbf{Abstract}\\
\end{center}
Droplet microfluidics are a powerful approach for hydrogel cell encapsulations. Much of the field has focused on single-cell encapsulations with pico-nanoliter droplet volumes necessary for single-cell sequencing or high-throughput screening. These small volumes, however, limit the use of hydrogel droplets for tissue engineering or cell therapies. We describe simple droplet microfluidics to generate millimeter-sized alginate droplets and demonstrate their use for cell encapsulations. This effort builds on our recent efforts \cite{arnold2023microfluidics}, specifically by replacing the glass slide forming the bottom layer of the chamber with a more hydrophobic acrylic (PMMA) layer to improve the alginate-in-oil droplet formation. Using glass layer and PMMA layer devices, we characterized the tunable production of water-in-oil droplets (average droplet lengths ranged from  0.8 to 3.7 mm). Next, PMMA layer devices were used to demonstrate the tunable generation of alginate-in-oil droplets (average droplet lengths ranged from 3-6 mm). Increasing the flow ratio (Q.ratio = Q.oil/Q.alginate) led to more uniform droplets as measured by the coefficient of variance (COV\%), which was approximately 5\%. Finally, a proof-of-use experiment used HUVEC-encapsulated alginate droplets as part of a scratch-healing assay. Specifically, HUVEC-encapsulated droplets (AH droplets) led to recovery of 3T3 fibroblast monolayers compared to no droplets or cell-free droplets (A droplets). Our results extended the use of simple microfluidics to generate and retrieve millimeter-sized alginate droplets for effective cell encapsulations. 
\end{sciabstract}

% Make a new page for the paper
\newpage
% Start introduction
\section{Introduction}
Hydrogel encapsulation of cells is an attractive approach for the sustained release of bioactive agents for tissue engineering in the laboratory and clinic \cite{lee2001hydrogels, correa2021translational}. Hydrogels provide 3D microenvironments supporting donor cells and prevent potential immune rejection. Droplet microfluidics has emerged as a valuable tool for hydrogel encapsulation of cells  \cite {white2021evaluation,chen2022trends}. Yet, driven by single-cell sequencing and other high throughput screenings, most droplets have focused on picoliter to nanoliter volumes \cite{klein2015droplet,guo2012droplet,agresti2010ultrahigh,vallejo2019fluorescence,price2016discovery}. These pL-nL volumes are essential for accelerating reactions or enabling sorting strategies. However, the limited reagent volumes (e.g., media for cells), preclude long-term cell culture applications via droplets due to waste buildup or depletion of reagents. To address this limitation, microfluidics or off-the-shelf droplet generators have been demonstrated to produce organoid and other 3D cultures \cite {trivedi2010modular,dolega2015controlled,zhang2021microfluidic}. However,  most existing systems maintain the generated droplets inside the tubing (in-situ) for long-term monitoring or provide limited opportunities to modify device properties \cite{zhang2021microfluidic,trivedi2010modular}. Therefore, developing alternative strategies to produce, modify device geometries, and retrieve millimeter-sized droplets for embedding cells via hydrogels is of great interest.\\

Various hydrogels (e.g., alginate \cite{tan2007monodisperse, zhang2021preparation}, Matrigel\cite{zhang2021microfluidic}, and polyethylene glycol diacrylate\cite{white2021evaluation}) have been used for droplet cell encapsulations \cite{rojek2022microfluidic}. In this effort, we used alginate, due to its robust and tunable mechanical properties \cite{drury2004tensile}, common use in tissue engineering, and broad availability. Previously Trivedi et al. \cite{trivedi2010modular}, using off-the-shelf components, had successfully demonstrated the generation of large cell-encapsulate alginate capsules/droplets that were up to 1.5 mm long. However, alginate capsules were permanently stored inside the tubing and were not retrieved. In another study, Zhang et al. used simple microfluidics combined with tubing to generate uniform Matrigel organoids up to 500 $\mu$m diameter)\cite{zhang2021microfluidic}. Here the organoids were solidified inside the tubing (temperature gelatin at 37$^{\circ}$C) and then collected for further investigations. Informed by these efforts, we sought to extend the use of mm-sized droplets by generating and retrieving alginate droplets. Additionally, we selected Human umbilical vein endothelial cells (HUVECs) as our target cells for encapsulation. Since HUVECs have been effectively use as a model for human endothelium and angiogenesis studies\cite{ rhim1998human, medina2020use}.\\

We demonstrated millimeter-sized HUVECs-encapsulated alginate droplets (Figure 1 A). To this aim, first, our recent co-flow \cite{arnold2023microfluidics} microfluidics were updated by replacing the glass slides forming their bottom layer with the more hydrophobic acrylic (PMMA) slides. Using water-in-oil experiments, glass-slide and PMMA-slide devices were characterized, and tunable droplet generations were verified (average droplet lengths ranged from  0.8 to 3.7 mm). Next, the PMMA-slide chips were used to generate tunable alginate-in-oil (average droplet lengths ranged from 3-6 mm). Increasing the flow ratios led to more uniform droplets as measured by the coefficient of variance (COV\%), which was approximately 5\%. Finally, HUVEC-encapsulated alginate droplets were generated and used as part of a scratch-healing assay. Specifically, HUVEC-encapsulated droplets led to faster healing of 3T3 fibroblast monolayers compared to no droplets or cell-free droplets. Collectively, these results extended the use of co-flow microfluidics to generate and retrieve millimeter-sized alginate droplets for effective cell encapsulations. 

\section{Materials and Methods}

\subsection{Microfluidics design and fabrication} 

Co-flow microfluidics were designed similarly to our recent effort \cite{arnold2023microfluidics}. Mold for the channels was designed with 2D CAD software. Figure 1. C demonstrates the design and critical dimensions of the channels. Briefly, the channels were 2 mm wide and 55 mm long. The diameter of the inlet was designed at 3.6 mm, and the outlet was 5 mm (Figure 1 C). Rectangular PMMA slides (68mm x 25mm) were also cut for experiments requiring a PMMA slide. These channels and the slides were cut from PMMA sheets (1/16" height, McMaster-Carr part number 8560K171) via a 40 W desktop CO$_{2}$ laser cutter (Glowforge Plus, Glowforge, Seattle, WA) following the manufacturer's recommendations for power and speed. 

After laser cutting, PMMA parts were retrieved and glued to a glass microscope slide (68 x 25 mm) using a strong adhesive (Super Glue, Gorilla) to create the fluidic molds. Molds were rested for approximately 30 minutes to allow for strong adhesion. 77 g of 
Polydimethylsiloxane (PDMS) (Dow Sylgard™ 184 Silicone 0.5 Kg Elastomer Kit) was prepared using the standard 10:1 w/w ratio (i.e., 70 g base and 7 g curing agent). Molds were placed inside a petri dish, and the PDMS mixture was added. Molds were placed inside a vacuum chamber (60 minutes or until clear of bubbles) to remove dissolved air bubbles. PDMS and the molds were baked for approximately 10 hours at 90$^{\circ}$C. 
Once the PDMS was fully cured, the microchannels were cut out with a border frame (15mm x 60mm) and separated from the remaining PDMS and glass slide. Care was taken to prevent detachment between the plastic mold and the glass slide. A first hole was made in the center of the inlet using a 2.5 mm diameter biopsy punch. A second 2.5 mm hole was cut inside the channel, 15 mm from the first hole. The first inlet was used for the continuous phase (oil), while the second was for the dispersed phase (water or alginate solution). The outlet was cut using a 5 mm biopsy punch. The larger outlet was helpful in maintaining droplet shape as the droplets exited the microfluidic channel. PDMS parts were permanently attached to glass slides via a standard plasma treatment (Harrick Plasma, Basic Plasma Cleaner, maximum RF power 18W) for the glass-slide systems. For the PMMA-slide devices, PDMS parts were attached using uncured PDMS. The finalized chip was cured at 90 $^{\circ}$ C for a minimum of 3 hours to ensure robust attachment. Care was taken to prevent any PDMS from coating the channel surfaces. Figure 1 D demonstrates the simplified workflow of building the PMMA-slide chambers.

\subsection{Cell culture and alginate-encapsulation} 
3T3 fibroblasts were obtained from ATCC (American Type Culture Collection, USA) and cultured in  DMEM supplemented with calf bovine serum. Human umbilical vein endothelial cells (HUVECs) were obtained from Lonza and cultured in EBM-2 media, supplemented with 10\% fetal bovine serum (Lonza) according to the provider's instructions. Cells were maintained at 37 $^{\circ}$C, 5\% CO$_{2}$, and the culture media was changed every 2-3 days. HUVECs from passages 5 to 12 were used in this study. Before cell-encapsulation experiments, HUVECs were collected by standard trypsin-EDTA treatment and suspended in the hydrogel culture and media. A 2\% (w/v) Sodium alginate solution was prepared by dissolving Sodium alginate powder (alginic acid sodium salt, Sigma Aldrich) in 1X Dulbecco's phosphate-buffered saline (1X PBS). 
HUVECs were seeded within these 2\% alginate solutions and cultured for 2 days at 10000 cells per mL density. A homogeneous hydrogel precursor solution containing 2\% (w/v) alginate and HUVECs were used as the dispersed phase for cell encapsulation. Food-grade olive oil containing 1\% (v/v) surfactant (SPAN 80) was used for the continuous phase. For water-in-oil characterization experiments, DI water with trace food dye was used. 

Cell and alginate (1 mL syringe) and oil (10 mL syringe) were each loaded  immediately before an experiment. Two independent syringe pumps (NE-300 Just Infusion - New Era) were used to deliver and control each solution. Silicone tubes (inner diameter of 1.5 mm) were used to connect each syringe (via a dispensing needle) to the chip. In cell-encapsulation experiments, the flow rate of the cell solution was kept constant at 0.5 mL/hr to avoid cell damage (e.g., due to shearing force during the injection). The flow rate of the continuous phase was set to 10 mL/hr. Outlet tubes were placed inside a graduated cylinder containing a 10 wt\% CaCl$_{2}$ solution. After the microfluidic production of the alginate droplets, droplets were delivered to the CaCl$_{2}$ solution. Droplets were incubated in the solution for 20 minutes before their collection for wound scratch assays. The system was housed inside a biosafety cabinet to prevent contamination. 

\subsection{Imaging and data analysis } 
For characterization experiments, the operation of the devices was recorded using a standard cellphone camera held via a tripod (UBeesize 10" Ring Light with Tripod Stand). Droplet dimensions were measured using NIH ImageJ software. In some experiments, since the droplets were elongated, the ratio of length over width was considered instead of the diameter. An EVOS Cell-Imaging system (bright-field images) was used for wound healing assays.

\section{Results}

\subsection{Characterizing PMMA-slide channels}
Droplet generation is informed by the channel's surface proprieties (e.g., hydrophobicity) added to considerations such as flow rates and the shape of the chamber. Most microfluidics use microscope glass slides as their base. Glass  has ideal imaging properties and provides sufficent mechanical durability. However, for aqueous-in-oil droplets (particularly hydrogels) the glass surfaces may need to be more hydrophobic. 
To this end, chemical treatments such as silanization or rain repellent have been used. These chemical treatments may need to be reapplied or could introduce other complications during fabrication. Moreover, for hydrogels, other rheological properties, for example, higher viscosity (e.g., alginate has a viscosity of 157 mPa.s compared to 1 mPa.s of water) could further complicate droplet formations. Therefore,  improvement to the system for increasing the channel hydrophobicity are desirable.  We considered replacing the standard glass layer with a more hydrophobic surface. PMMA and PDMS (both more hydrophobic than glass) were considered for this aim. While PDMS is more hydrophobic than PMMA, it was determined that attaching the PDMS channels to PMMA slides were more reliable. PMMA slides were also more mechanically durable as compared to PDMS. To verify the hydrophobic properties of PMMA, contact angles for water and 1\% alginate solutions on both glass and PMMA were measured (supplemental section). Experiments confirmed that both solutions had a higher contact angle on PMMA as compared to glass. It was hypothesized that the PMMA chips, due to their hydrophobicity, would produce more uniform droplets as compared to the standard chips. To test this hypothesis first, water-in-oil experiments were conducted. \\

In this report, PDMS chambers with the exact dimensions were made. The first chamber was attached to a glass slide via plasma treatment.
The second chamber was attached to a PMMA slide. Both chambers were tested under the same volumetric flow conditions to determine the role of surfaces in droplet formation. Specifically, four flow ratios (Q.ratio = Q.oil/Q.water) were tested for each chip. These were Q.ratio 2, 3, 4, and 5. For these experiments, the volumetric flow rate of oil was fixed (Q.oil = 5 mL/hr), and the Q.water was adjusted. These flow rates were selected based on their success in our preliminary experiments. 
After the successful generation of water-in-oil droplets, tunable generation of alginate droplets were investigated. Alginate is a linear polyanionic block copolymer composed of two units; its gelation is mediated through cations - typically calcium ions. The glass layer devices resulted in threading or merging droplets. Therefore their use were abandoned. Here the results from PMMA-slide devices are shared. Figure 3 demonstrates the tunable production of droplets for four Q.ratios of 2, 3, 4, and 5. The volumetric flow rate of the oil was fixed (Q.oil = 5 mL/hr). In this experiment, each flow condition produced a statistically different droplet size. The length/width ratio ranged from 2-4 for the alginate droplets. Histogram of droplet size and the coefficient of variance (COV\%) for each condition is reported. 
After the successful generation of water-in-oil droplets, tunable generation of alginate droplets were investigated. Alginate is a linear polyanionic block copolymer composed of two units; its gelation is mediated through cations - typically calcium ions. The glass layer devices resulted in threading or merging droplets. Therefore their use were abandoned. Here the results from PMMA-slide devices are shared. Figure 3 demonstrates the tunable production of droplets for four Q.ratios of 2, 3, 4, and 5. The volumetric flow rate of the oil was fixed (Q.oil = 5 mL/hr). In this experiment, each flow condition produced a statistically different droplet size. The length/width ratio ranged from 2-4 for the alginate droplets. Histogram of droplet size and the coefficient of variance (COV\%) for each condition is reported. 
Droplet generation was recorded, and ImageJ was used to measure the properties of all droplets. For each droplet, specifically, length and width were measured. The length-to-width ratio was considered as the critical metric to characterize each droplet. Figure 2 summarizes the results of these experiments. As expected, increasing the Q.ratio led to smaller droplets. Channels with PMMA layers generated more uniform droplets as measured by their standard deviation. For both systems the length over width ratio ranged from 1 - 3. Histogram of droplet properties  and the coefficient of variance (COV\%) for each condition is reported. The frequency of droplet generation was similar for both devices.

\subsection{Tunable generation of alignate droplets}

After the successful generation of water-in-oil droplets, tunable generation of alginate droplets were investigated. Alginate is a linear polyanionic block copolymer composed of two units; its gelation is mediated through cations - typically calcium ions. The glass layer devices resulted in threading or merging droplets. Therefore their use were abandoned. Here the results from PMMA-slide devices are shared. Figure 3 demonstrates the tunable production of droplets for four Q.ratios of 2, 3, 4, and 5. The volumetric flow rate of the oil was fixed (Q.oil = 5 mL/hr). In this experiment, each flow condition produced a statistically different droplet size. The length/width ratio ranged from 2-4 for the alginate droplets. Histogram of droplet size and the coefficient of variance (COV\%) for each condition is reported. 

\subsection{Wound-scratch Assay}
Next, we encapsulated HUVECs via uniform droplets and demonstrated their use as proof-of-concept cell therapeutics. HUVECs were cultured and pre-mixed with alginate, as described in the methods section. The alginate-cell solution and the oil were loaded syringes immediately before the experiments. Syringes were attached to a chip via 1.5 mm silicone tubing. Care was taken to prevent bubbles inside the tubes. The chips were primed with oil before the experiments. The flow rate for the cell solution was set to 0.5 mL/hr, and the flow rate of the oil was set to 10 mL/hr (leading to Q.ratio = 20). Alginate droplets generated inside the chip were delivered to an 20 mL graduated cylinder via tubing containing a 10 wt\% CaCl${_2}$ solution for polymerization. The droplets remained inside the solution for 20 minutes before their removal for the wound-scratch assay. 

For the wound scratch assay, monolayers of 3T3 fibroblast cells were prepared. Cells were seeded on a 12 well-plate at a density of 6500 cells per cm${^2}$. A 10 $\mu$L pipette tip was used to generate the wound to remove the cells from the monolayers. Cells were washed with PBS to remove any floating/non-adhering cells. Using a marker, additionally, the positions of the wounds were marked on the bottom of the 12 well-plate. The following conditions were considered to determine the efficacy of the alginate-encapsulated HUVEC droplets (AH droplets). The first control, only the monolayer (with no droplet) was used. The second control used alginate droplets (A droplets) with no HUVECs. Finally, for the experimental condition, AH droplets were used. The initial gaps for the wounds were imaged and quantified via ImageJ. For AH and A droplets, four droplets were placed at each corner of the wound on the monolayer. The wound healing was assessed by measuring the gap after two days. Moreover, to control for the effects of media, experiments were conducted for both the 3T3 media and the HUVEC media. Figure 4 provides the results of the experiments. 

For both 3T3 and HUVEC media conditions, the initial gaps for No droplets (first control) were reduced. 3T3 cells grew better in the 3T3 media and reduced the gap more when  compared to HUVEC media.  For both media conditions, the gap decreased in the presence of the AH droplets for both types of media. The A beads (no HUVECs) with HUVEC media did not result in a statistically significant change in the gap. The A beads with the 3T3 media increased the gap. Collectively, these results provided a proof-of-concept application of cell-encapsulated alginate beads in the context of a standard scratch assay and wound healing. Experiments were again replicated, and gaps were measured after 1 days. Similar results were observed. 

\section{Discussion}
We investigated the use of co-flow microfluidics to generate millimeter-sized alginate droplets. Applying the hydrophobic PMMA-layer was essential for this goal, specifically compared to the standard glass-layer devices. We had previously used co-flow microfluidics to generate millimeter-sized water-in-oil and Matrigel-in-oil droplets \cite{arnold2023microfluidics}. We selected alginate and HUVECs specifically for cell-therapeutics due to the well-established use of alginate in wound healing, tissue engineering \cite{aderibigbe2018alginate,zhang2020alginate},  and versatile application of HUVECs in tissue engineering \cite{medina2020use}. At the start of this investigation, glass-layered co-flow microfluidics was ineffective in the uniform production of alginate-in-oil droplets. Various issues (threading, non-uniform productions, and alginate droplets merging) were observed. To address these issues, we sought to increase the hydrophobic properties of the channel. Therefore, we replaced the glass layer with PMMA. The PMMA layer enabled the robust production of large alginate droplets. Unlike other approaches that increase the hydrophobic properties, for example, via a surface coating, there was no need to reapply the treatment. The PMMA approach was selected due to its easy fabrication and robust mechanical durability. However, if an increase in hydrophobic properties were needed, other materials (including PDMS) can be explored. 

Previous studies have successfully used off-the-shelf components and capillary glass tubes to establish droplet generators \cite{gu2011theoretical}. In one great example, Trivedi et al. demonstrated a robust and practical approach for large droplets (up to 1.5 mm) using off-the-shelf components such as tubing and connectors \cite{trivedi2010modular}. Moreover, they were able to generate alginate droplets and in-situ culture of cells (up to multiple days) inside the Teflon (PTFE) tubes. This approach is practical and inexpensive for the production and storage of droplets. However, as the paper describes, realizing complex geometries with the system may be difficult (or often impossible). Additionally, the size of the droplets depend on a tube's dimensions (inner diameter) and their availability. The use of additive manufacturing (3D printing) was also considered. Indeed, 3D printing approaches are highly suitable for realizing complex geometries. While significant progress has been made in addressing potential challenges\cite{musgrove2022applied}, some concerns remain regarding the biocompatibility and potential toxicity of 3D printed parts (particularly the more desirable STL systems) \cite{oskui2016assessing}. We sought to avoid these complications by using desktop laser cutters. Similar to 3D printers, in a university setting, laser cutters are broadly available in university (and even community) maker spaces.   

\section{Conclusion}
The use of droplet microfluidics for single-cell assays, diagnostics, high-throughput screening, and liquid handling systems has increased over the past decade. Yet the typical small pL-nL volumes preclude the application of the droplets for long-term cell cultures. Recent advances have been made to generate mm-sized and in-situ cell cultures of organoids and capsules. However, the ability to retrieve these cultures from the device/tube and potential modifications to the channel geometries are highly desirable. Specifically, the shape of the chambers and nozzles may be essential in scaling and controlling the robust production of these larger droplets. The approach and outcomes presented here provide a simple and clean-room free for large droplet generation and seek to broaden access to these useful technologies. The alginate-encapsulated HUVEC droplets system has obtained good activity to promote wound healing by enhancing angiogenesis and cell migration. This system could incorporate bioactive agents that facilitate wound healing and promote angiogenesis (e.g., VEGF). Alginate droplets could also be used for drug delivery vehicles and therapeutic development. This research provides a novel and versatile strategy for fabricating cellularized hydrogel constructs with clinically relevant dimensions.

\newpage
\section* {Acknowledgments}  
Authors express gratitude to members of Ahrar-Lab (specifically Brian Le).\\
This work was supported in part by CSULB startup funds and a CSUPERB new investigator award to SA. Additionally, this work was supported by National Science Foundation grant to PA.\\
Funding agencies are not responsible for the content of this manuscript.\\
KT was supported by the CSULB’s NIH BUILD Scholars program.\\

\section* {Conflicts of Interest} 
The authors declare no conflict of interest.

\section* {Additional Resources}

\section*{Corresponding Authors Addresses}
Siavash Ahrar (Ph.D.)\\
Mail: Department of Biomedical Engineering, CSU Long Beach, 
1250 Bellflower Blvd. \\VEC-404.A, Long Beach, CA 90840, 
E-mail: Siavash.ahrar@csulb.edu\\

Perla Ayala (Ph.D.)\\
Mail: Department of Biomedical Engineering, CSU Long Beach, 1250 Bellflower Blvd.\\
ET-108, Long Beach, CA 90840 E-mail: Perla.ayala@csulb.edu\\

\textbf {List of Figures}:
\begin{itemize}  
  \item Figure-1: Millimeter-sized Alginate Droplet generators. 
  \item Figure-2: Water-in-oil droplet production with PMMA-layer and Glass-layer chambers.
  \item Figure-3: Alginate-in-oil droplets with PMMA-layer devices.
  \item Figure-4: The role of Alginate-encapsulated HUVECs droplets (AH Droplets) in Wound healing (scratch assay). 
\end{itemize}

\textbf {List of Supplemental Figures}:
\begin{itemize}  
  \item S.Figure-1: Sample contact angles. 
  \item S.Figure-2: Alginate-in-oil droplets with glass-layer devices demonstrating Alginate threading. 
\end{itemize}

\newpage
\section{Supplemental Materials}

\subsection{Contact-angle characterization}
Contact angle measurements were carried out with water and alginate to determine the hydrophobicity and surface properties of untreated glass and PMMA. Droplets were placed on the solid sample of Glass or PMMA. A high-resolution camera of CellScale and software IC capture 2.5 were used to capture the image of the droplets. ImageJ was used to calculate the angles.

\subsection{Code and design availability}
Analysis code developed for this manuscript is available upon request.

\subsection{Statistical analysis}
Statistical analysis and plots for the experiments were conducted via R. 
In the analysis, the following p-value were considered:\\
ns: 5.00 $\times 10^ {- 2}$ < p <= 1.00 \\
*: 1.00$\times 10^ {- 2}$< p <= 5.00$\times 10^ {- 2}$\\
**: 1.00$\times 10^ {- 3}$ < p <= 1.00$\times 10^ {- 2}$\\
***: 1.00$\times 10^ {- 4}$ < p <= 1.00$\times 10^ {- 3}$\\
****: p <= 1.00$\times 10^ {- 4}$ 

\newpage
\begin{flushleft}
\bibliographystyle{unsrt}
\bibliography{refs} % Entries are in the "refs.bib" file
\end{flushleft}

\newpage

\begin{figure}[b]
\includegraphics[width=\textwidth]{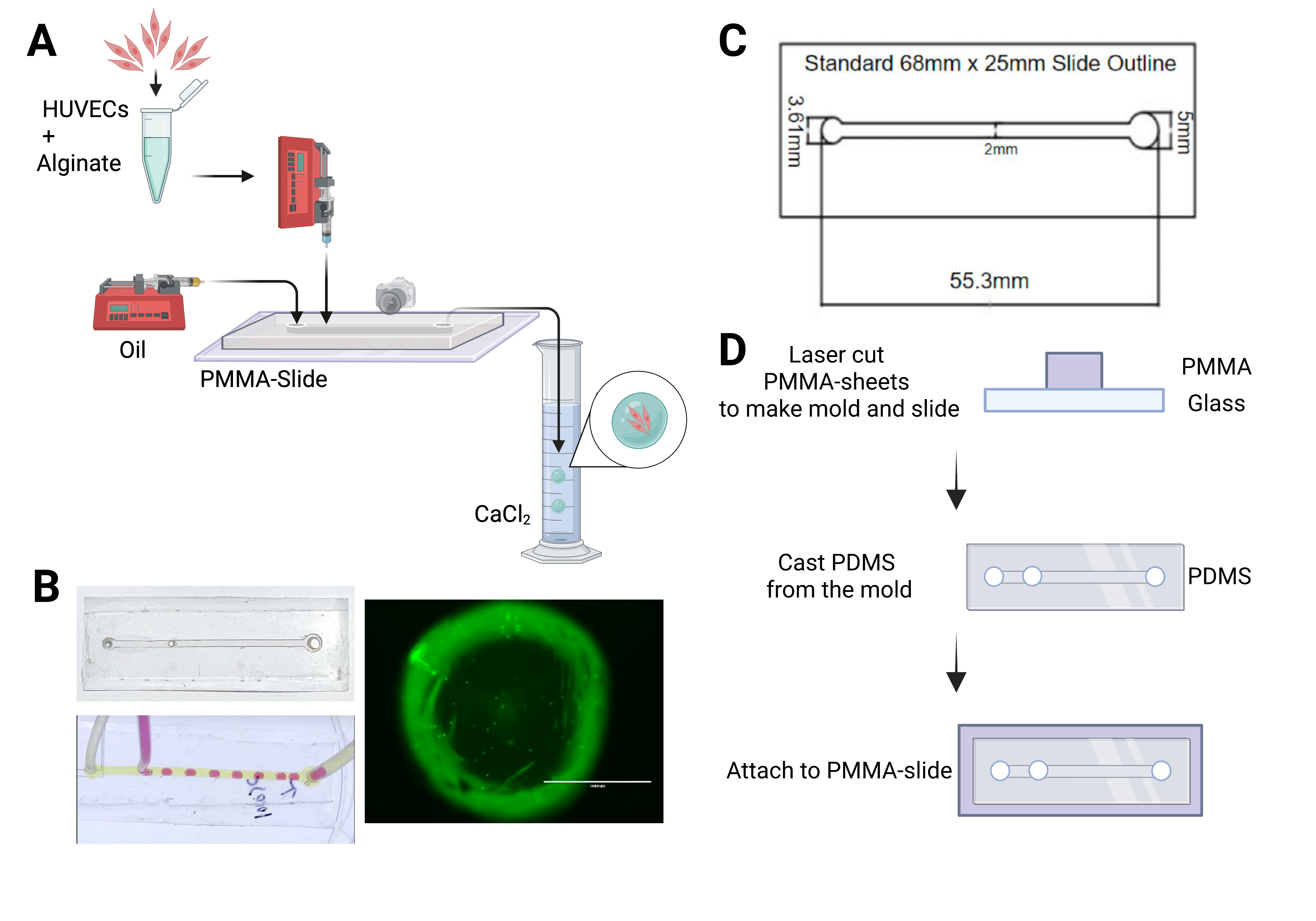}
\centering
\caption{\textbf {Millimeter-sized Alginate Droplets}. (A) Co-flow microfluidics for the encapsulation of cells (HUVECs) inside millimeter-sized alginate droplets. Cell were pre-mixed with alginate. Alginate-in-oil droplets are generated and can be tuned. The droplets were polymerized inside the CaCl${_2}$ solutions. (B) Image of the chip, production of water-in-oil droplets, and example of HUVECs encapsulated droplets with cells stained. (C) Key dimensions and layout for the droplet generator chip. (D) The workflow to manufacturing the mold and the microfluidics device with a PMMA layer. Figure in part produced by BioRender.}
\end{figure}

\begin{figure}[b]
\includegraphics[width=\textwidth]{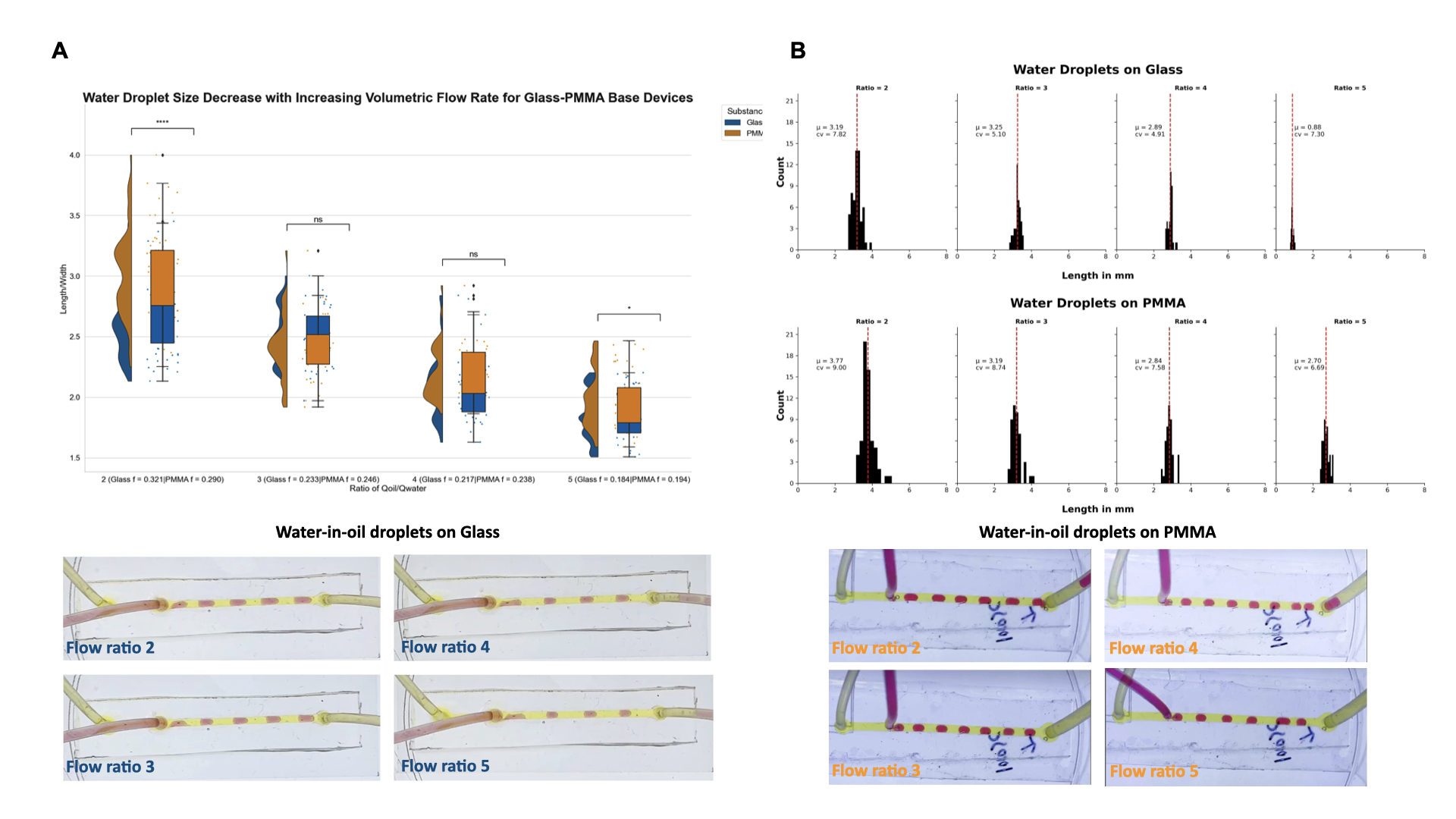}
\centering
\caption{\textbf {Water-in-oil droplet production with Glass-layer and PMMA-layer devices}. (A) Tunable generation of water-in-oil droplets via glass-layer and PMMA-layer devices. Increasing the flow ratio led to smaller droplets. Here droplet size is characterized via the ratio of droplet length to width. (B) The histogram of droplet length for Glass and PMMA layer devices. (C) Water-in-oil droplets generated via glass-layer chambers. Water-in-oil droplets generated via PMMA-layer chambers Examples of droplet generators. (Flow ratios 2,3,4, and 5).}
\end{figure}

\begin{figure}[b]
\includegraphics[width=\textwidth]{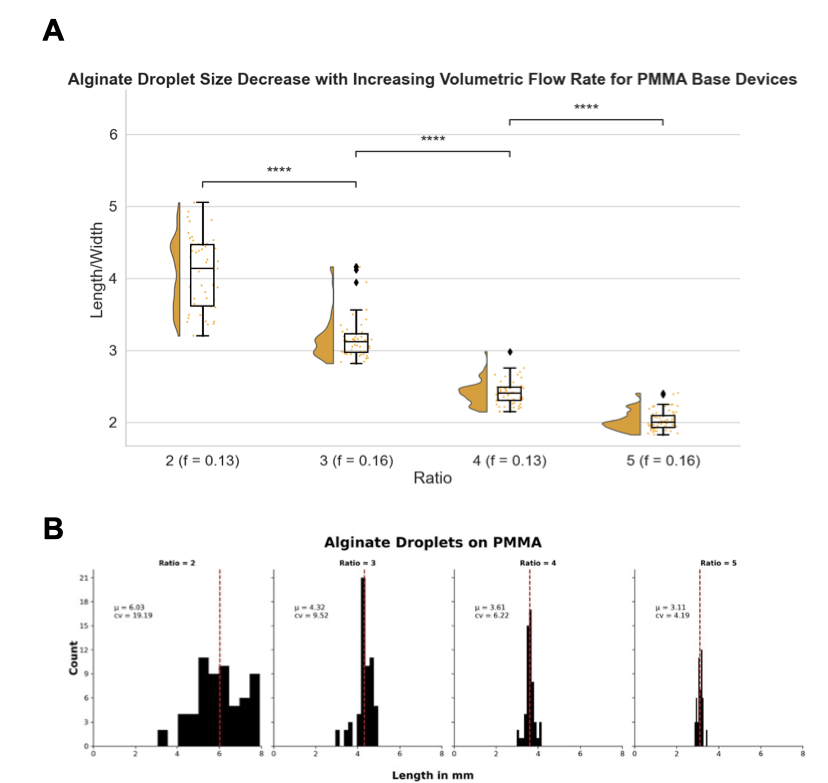}
\centering
\caption{\textbf {Alginate-in-oil droplet production with PMMA-layer devices}. (A) Tunable generation of alginate-in-oil droplets via PMMA-layer devices. Increasing the flow ratio led to smaller droplets. The droplet size is characterized via the ratio of droplet length to width (B) The histogram of droplet length for  PMMA layer devices. Increasing the Flow ratio leds to more uniform droplets.}
\end{figure}

\begin{figure}[b]
\includegraphics[width=\textwidth]{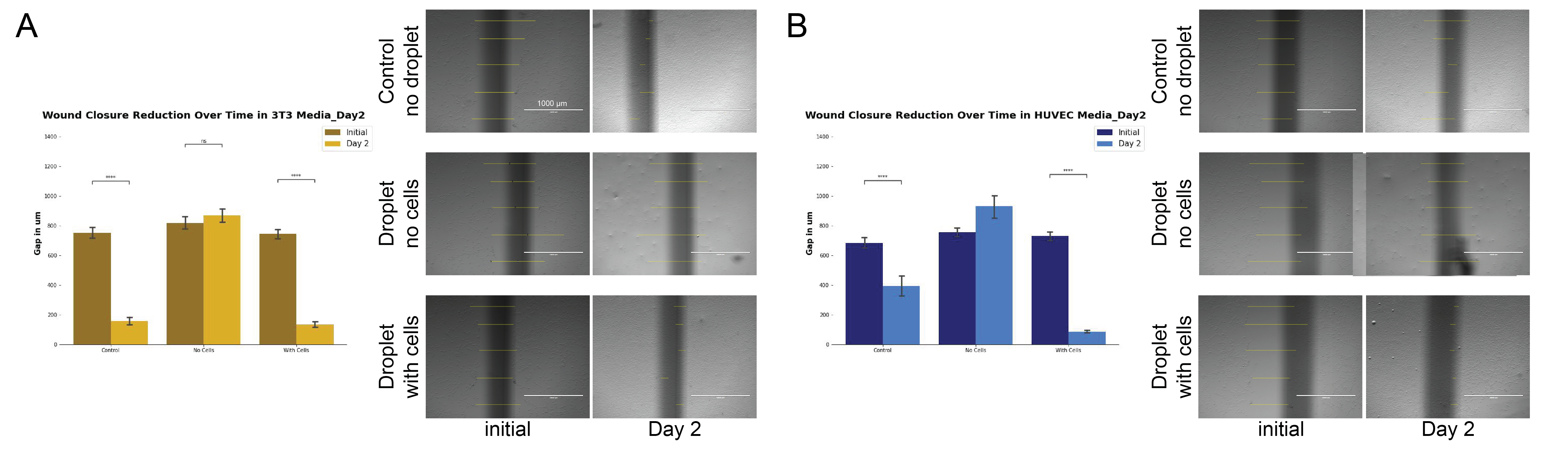}
\centering
\caption{\textbf {Demonstration of the role of alginate-encapsulated HUVECs droplets in wound healing via a scratch assay}. (A) Characterization of the wound gap initial day and day 2 with no droplets, alginate droplets with no cell, and HUVECs encapsulated droplets when cells were cultured with 3T3 media. Droplets with cells and no droplet conditions led to wound healing. The gap increased for alginate droplets without cells. Yellow lines indicate distances that were used from sample images to quantify gaps. (B) Characterization of the wound gap initial day and day 2 with no droplets, alginate droplets with no cell, and HUVECs encapsulated droplets when cells were cultured with HUVEC media. Similar to the previous experiments, droplets with cells and no droplet conditions led to wound healing; gap increased for alginate droplets without cells. Sample images for each condition demonstrating measurments between boundaries/gaps at the start of the experiment and day 2.}
\end{figure}

\begin{figure}[b]
\includegraphics[width=\textwidth]{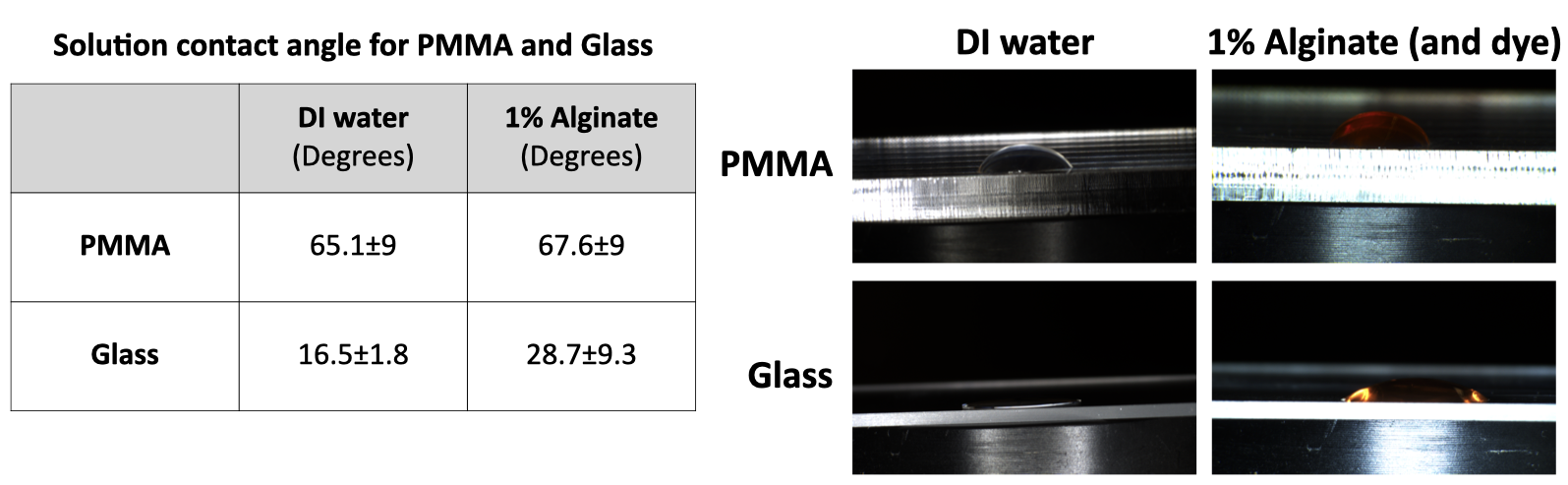}
\centering
\caption{\textbf {Supplemental Materials Figure-1: Contact Angles}. (A) Table summary of contact angle measurements. PMMA is more hydrophobic than glass. (B) Sample images of DI water and alginate solutions on glass and PMAA. }
\end{figure}

\begin{figure}[b]
\includegraphics[width=\textwidth]{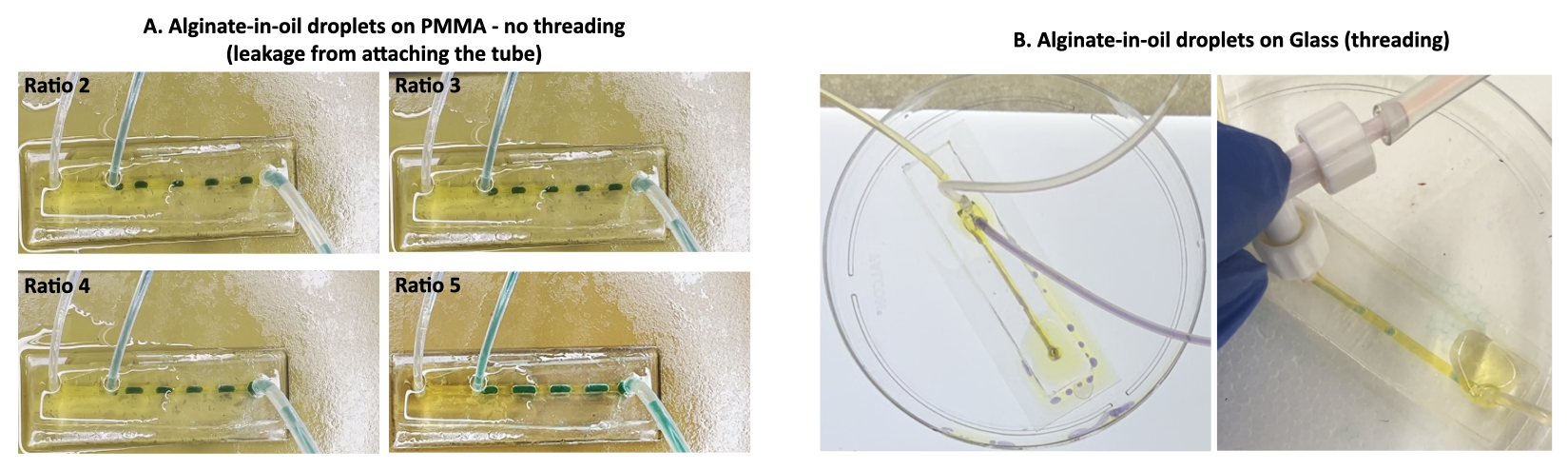}
\centering
\caption{\textbf {Supplemental Materials Figure-2: Alginate-in-oil droplet PMMA vs Glass}. (A) Alginate-in-oil droplets on PMMA layer devices (flow rations 2,3,4, and 5). Droplets form on PMMA surfaces without threading. (B) Glass layer devices led to threading. }
\end{figure}

\end{document}